\documentclass{article}

\usepackage{arxiv}

\usepackage[utf8]{inputenc} % allow utf-8 input
\usepackage[T1]{fontenc}    % use 8-bit T1 fonts
\usepackage{hyperref}       % hyperlinks
\usepackage{url}            % simple URL typesetting
\usepackage{booktabs}       % professional-quality tables
\usepackage{amsfonts}       % blackboard math symbols
\usepackage{nicefrac}       % compact symbols for 1/2, etc.
\usepackage{microtype}      % microtypography
\usepackage{lipsum}		% Can be removed after putting your text content
\usepackage{graphicx}
\usepackage{natbib}
\usepackage{doi}
\usepackage[table,xcdraw]{xcolor}
 \usepackage[normalem]{ulem}
\usepackage[utf8]{inputenc}
\usepackage{tikz}
\def\checkmark{\tikz\fill[scale=0.4](0,.35) -- (.25,0) -- (1,.7) -- (.25,.15) -- cycle;}

\title{WildKey: A Privacy-Aware Keyboard Toolkit for Data Collection In-The-Wild}

\author{ {André Rodrigues} \\
	LASIGE, Faculdade de Ciências,\\
	Universidade de Lisboa\\
	\texttt{afrodrigues@fc.ul.pt} \\
	\And
	{André Santos} \\
	LASIGE, Faculdade de Ciências,\\
	Universidade de Lisboa\\
	\texttt{abranco@lasige.di.fc.ul.pt,} \\
		\And
	{Kyle Montague} \\
	NorthLab,\\
	University of Northumbria\\
	\texttt{kyle.montague@northumbria.ac.uk} \\
			\And
	{Hugo Nicolau} \\
	INESC-ID, Instituto Superior Técnico\\
	Universidade de Lisboa\\
	\texttt{hman@inesc-id.pt} \\
		\And
	{Tiago Guerreiro} \\
	LASIGE, Faculdade de Ciências,\\
	Universidade de Lisboa\\
	\texttt{tjvg@di.fc.ul.pt} \\
}

\renewcommand{\headeright}{A Working Paper}
\renewcommand{\undertitle}{A Working Paper}
\renewcommand{\shorttitle}{WildKey}

\hypersetup{
pdftitle={WildKey},
pdfsubject={},
pdfauthor={André Rodrigues},
pdfkeywords={Text-Entry, Touch Dynamics, In-the-wild, Smartphones, Data Collection},
}

\begin{document}
\maketitle

\begin{abstract}
Touch data, and in particular text-entry data, has been mostly collected in the laboratory, under controlled conditions. While touch and text-entry data has consistently shown its potential for monitoring and detecting a variety of conditions and impairments, its deployment in-the-wild remains a challenge. In this paper, we present WildKey, an Android keyboard toolkit that allows for the usable deployment of in-the-wild user studies. WildKey is able to analyse text-entry behaviours through implicit and explicit text-entry data collection while ensuring user privacy. We detail each of the WildKey’s components and features, all of the metrics collected, and discuss the steps taken to ensure user privacy and promote compliance. 
\end{abstract}

% keywords can be removed

\section{Introduction}
The widespread use of smartphones has led to a massive increase in the generation of personal data, with text input being one of the most common tasks. People type to send messages, write emails, engage in social networks and much more. The data generated from these interactions has remarkable potential as a digital endpoint for disease detection and monitoring, to the quantified self movement, and biometrics. As a digital endpoint, text-entry metrics have been used to distinguish between people with Parkinson’s Disease (PD) and control groups showing potential for early disease detection \cite{gallego2017,Iakovakis2018,dhir-etal-2020-identifying}, to assess stress \cite{ciman2015}, fatigue \cite{al-libawy2020}, distinguish between patients with multiple sclerosis and controls \cite{lam2020}, and even for ubiquitous inebriation assessment \cite{Mariakakis2018}. In the field of user authentication, keystrokes dynamics have been used to discriminate among users detecting potential impostors, indicating that typing behaviour is highly personal \cite{Killourhy2009}. Preliminary work also suggests that differences between typing sessions can be associated with users’ emotions \cite{Hadjidimitriou2017}. Despite all the potential shown by recent work, collecting this type of data, and particularly in-the-wild, raises several challenges in regards to required user effort, compliance, privacy, ease of deployment and study oversight. 

There have been multiple approaches to collecting typing data in-the-wild. The most simple is explicitly prompting the user to do a specific task (e.g. transcription tasks) in a custom made application \cite{Reyal2015}. However this requires a lot of effort from the user, is not able to collect natural typing behaviour, nor provide spontaneous assessments,  and overall one can expect less compliance as more data is requested from users. It has the benefit of mimicking controlled laboratory tasks which may provide data with less noise. A second approach is to collect everything the user is typing by logging keypresses and touches \cite{nicolau2017}, which faces issues regarding user privacy and overall adherence to the study. A third solution is to only collect typing metrics that are not related with the content written such as flight time and hold times \cite{Iakovakis2018}. The approach ensures user privacy at the cost of limiting the type of metrics one can extract from typing sessions. A similar approach is to obfuscate the text written, either by only parsing parts of the data \cite{Buschek2018} or by analysing the typing session and only storing an abstract representation of the sentences (e.g. “Noun Verb Adjective” or substituting every letter with an "M") \cite{Bemmann2020, evans2012}. Depending on the method used, a different set of potential features  can be extracted. All of the approaches described have advantages and disadvantages to them, and depending on the context, different ones, or a combination of several may be the ideal solution.

The WildKey toolkit was designed to support the collection of data both explicitly with prompted tasks, and implicitly providing the best of both. With the WildKey keyboard one  can passively analyse all text written regardless of the application the user is in. Unlike prior approaches, WildKey neither stores raw textual data, nor it obfuscates typing sessions. Instead, for implicit data collecting we shifted the analyses that required access to the raw text to the device, and only calculated metrics with no potential to reconstruct the text are stored. The toolkit enables researchers and developers to tailor the WildKey data collecting to what best suits their study.  The WildKey keyboard supports an unconstrained text-entry protocol \cite{MacKenzie2001} where users are free to write however they like including using suggestions, auto-correct and cursor changes \cite{Zhang2019}. It is able to provide all the traditional text-entry metrics on speed and error rates \cite{MacKenzie2001} and touch dynamic data that has been key in multiple studies \cite{gallego2017,Iakovakis20188,lam2020} among others. As a toolkit for in-the-wild data collection it provides additional features that are commonly needed and even required to successfully run this type of studies. In addition to the text-entry tasks, one can create and schedule questionnaires and other custom made tasks such as the Alternate Finger Tapping assessment. Lastly, the toolkit has a Study Manager application that enables researchers/developers to easily schedule, deploy and oversee their active studies.

\section{WildKey}
\label{sec:headings}
The WildKey toolkit is composed of a \textit{Keyboard Android app},\textit{ a} \textit{Study Management \& Analytics app (React), and a NoSQL Database (Firebase)}. The toolkit was developed to enable researchers and developers to extend and deploy their own standalone ecosystems. The repository is open source, under the license Attribution 4.0 International (CC BY 4.0) and available at \url{https://techandpeople.github.io/keyboard/}.  

The \textbf{WildKey keyboard } extends the Android Open Source Project Keyboard \cite{AndroidO77:online}. As such, it supports 26 languages, provides auto-correct, suggestions, customization options among many other traditional keyboard features. WildKey keyboard is responsible for collecting, and processing all the data inserted and storing its results on the defined cloud database storage. The details of the types of data collection and tasks available is discussed below in the \textit{Data Collection}. A demo standalone version with local storage is available in the Playstore [toAppear]. 

The \textbf{Study Management \& Analytics} application allows researchers to define and schedule their user studies and associate study tasks/schedules to registered users (details below in Study Management). Through it, we also offer a variety of features to support overseeing studies and quick visualizations of text-entry related metrics (\textit{Metrics section}). 

The database is responsible for storing all the information collected from each individual user and it is where all study configurations and schedules are stored. The keyboard app relies on these configurations to load and schedule tasks to its users, providing notifications, and other nudges (details below under Study Management).

The WildKey toolkit was designed to be \textbf{Privacy-Aware} in order to not only comply with the current standards for data protection, but also promote user compliance and adherence to study protocols. As such, during implicit text-entry collection (i.e. when the user is writing using the keyboard in any and all text fields regardless of application) no raw text is ever recorded on device storage or sent to the cloud, nor any data that would enable anyone to reconstruct the text-written. All text-entries  that require content analysis are done locally on the device, and only processed data is sent to the cloud database. We highlight these and other design choices in the \textit{Privacy-Aware }section.

\subsection{Data Collection}
The WildKey Keyboard app is prepared to collect three types of data: text-entry, questionnaires and custom made tasks. For text-entry, the keyboard is able to assess \textit{Implicit Text-entry} data and to prompt users to do text-entry tasks collecting \textit{Explicit Text-Entry} data. The keyboard app also enables developers to create \textit{Questionnaires} to be answered within the app that accompanies the keyboard. Lastly, we designed the keyboard to be flexible and support the creation of \textit{Custom-Made} tasks relevant for a specific study protocol. 

All data collected is sent via hypertext transfer protocol secure (HTTPS) and by default is prepared to store data in a NonSQL database (Firebase) with JavaScript Object Notation (JSON). The toolkit has an isolated data controller which enables developers to easily integrate an alternative data controller to match the database backend desired (e.g. MySQL).

When users first install the WildKey keyboard app they will be asked to create a user account by registering using an email address. WildKey will create a user profile and store in the database only a tokenID representative of the user with the email being used only for the first authentication. Additionally, Wildkey will collect information about the user device (i.e. operating system version, brand, model, screen dimensions and resolution).

\begin{figure}
 \includegraphics[scale=0.73]{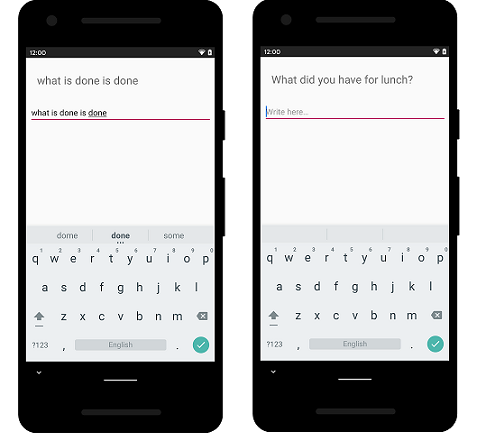}
 	\centering
  \caption{Transcription and Composition task example.}
  \label{fig:fig1}
\end{figure}

\subsubsection{Explicit Text-Entry Collection}
There are two types of text-entry tasks supported by WildKey. The first is the standard transcription task used in traditional text-entry studies (i.e. asking the user to transcribe a set of phrases). The phrases to be provided to each user can be randomized and are provided when establishing the study protocol within the \textit{Study Management}. Users are encouraged to type as accurately and quickly as possible. Transcription tasks allow us to know for certain what was the target intended phrase (since we provide it), thus enabling us to calculate error rate related metrics with high confidence. However, transcription tasks are not engaging for users, can be more demanding, do not portray natural typing behaviours and are not able to collect data spontaneously from users, nor for extended or frequent periods of time.

The second explicit text-entry task WildKey supports is Compositions. Users can be prompted at scheduled times to answer a sequence of questions following a similar protocol to the one described  above. Unlike the Transcriptions tasks we do not have the ability to know for certain what was the intended sentence, or sentences the user was trying to write. Thus, to calculate metrics related with error rates we relied on the approach presented in \cite{evans2012} where target intent is calculated using a spell checker. Word by word, if it exists we consider it to be the equal to the user intent. In instances where the word does not  exist we rely on the sentence written until the unknown written word (including it) and use the spell checker to predict the intended word, the top recommendation is chosen as the intended word to calculate all metrics. During the transcription tasks we also calculate and provide this calculated sentence based on the user input and spell checker to verify reliability. Composition tasks offer the possibility to ask questions relevant to the context of the user study providing additional key information (e.g. “What did you eat for lunch?”), are potentially more engaging for users, and typing behaviours will be more natural than ones triggered in transcriptions tasks. However, this type of task still faces similar challenges in its ability to collect data spontaneously, for extended or frequent periods of time.

WildKey supports an unconstrained text-entry protocol \cite{mackenzie_tanaka-ishii_2007}, where users are always free to correct any errors they encountered. Each text-entry trial (within an explicit task) has at the top of the screen the sentence/question to transcribe/answer, and an edit box below. The trial starts once the user writes the first letter, and ends when the user writes the last letter. The next trial is shown once the user confirms it on the keyboard. If it is the last one, users are sent to a completed task screen and shown when the next task is scheduled to. 

Exclusively during explicit text-entry collection tasks, WildKey collects raw text, all user touch point data and calculates detailed error metrics (i.e. Omissions, Substitutions and Insertions - discussed below in Analytics).

\subsubsection{Implicit Text-Entry Collection}
When the keyboard is installed and active, all text written, regardless of where it is written, is analysed to calculate all the metrics described below in Analytics. The exceptions to the rule is when users are writing in password fields or inserting only numbers, where no metrics are calculated.

When the user opens the keyboard in any field, WildKey considers the first key entered to be the start of a text-entry trial, and the trial is ended once the keyboard is hidden again and the last key entered represents when the trial stopped. Furthermore, if more than a set amount of time elapses between two key inserts, we segment the sessions and consider it to be a new session. This allows us to segment text-entry sessions and calculate metrics for each of these trials (e.g. words per minute). 

When the user opens the keyboard in a text edit field with text already present, and at some point during the trial it changes its cursor to the previously written text, we ignore the text-entry session and mark it as discarded. Since the text was written in a previous session we do not  have the ability to compute any metrics (without compromising its reliability) and therefore the session is discarded. Nevertheless WildKey registers that the trial has been discarded and records some high level metrics for the session such as number of characters written. 

Similar to composition tasks, during implicit data collection we do not know the user intended target sentence. As such we use the protocol described in the previous section to calculate user intent enabling us to assess error related metrics.

To preserve users' privacy and promote compliance, no text or any data that would enable its reconstruction is recorded during implicit collection. All metrics are calculated locally on the device and only metrics with no textual content are sent and stored in the cloud database. 

\subsubsection{Processing Text-Entry Data}
While the user is writing, WildKey is creating an input stream buffer with all the text inputs. Simultaneously, Wildkey stores an array of the actions performed, an array of all cursor changes, and a list array of all the suggestions given by the spell checker with each letter entered. The array of actions is composed of corrections (i.e. deletes or substitutions) and entry actions \cite{Zhang2019}. The cursor changes array allows Wildkey to adjust the input stream at the end of the text-entry trial to account for non sequential changes to the text. In explicit text-entry collection with transcription tasks we rely on the target phrase provided to the users to compute all error related metrics. For implicit and composition tasks, the suggestions array enables WildKey to predict the user intent based on the spell checker predictions of the intended word. The input stream is then processed locally by Wildkey to compute all the metrics described below in Analytics; all other information is discarded. The calculated metrics are synchronized to the cloud database when an internet connection is available. 

\begin{figure}
 \includegraphics[width=\textwidth]{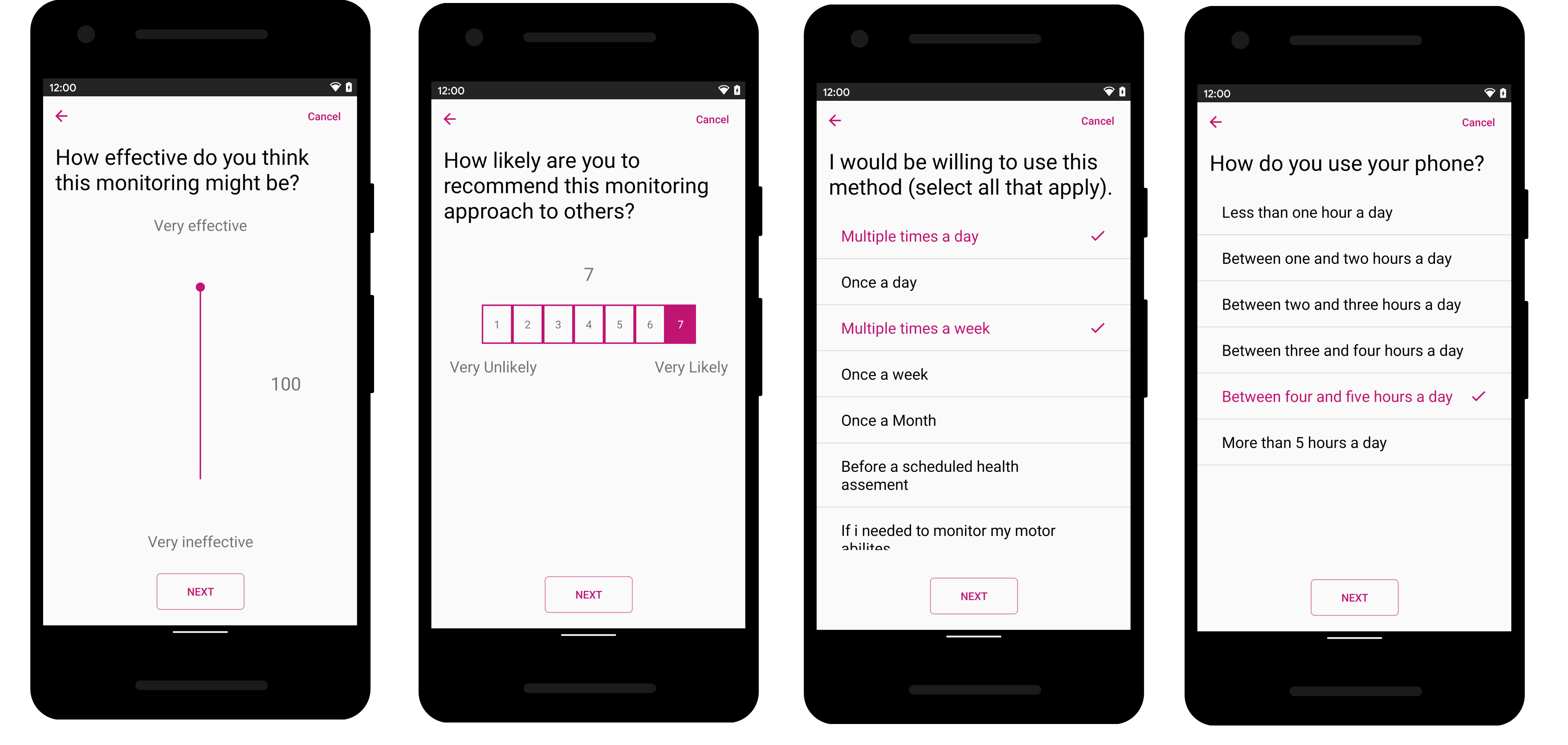}
 	\centering
  \caption{Four of the questions types available: slider scale, button scale, multiple choice and single choice.}
  \label{fig:fig2}
\end{figure}

\subsubsection{Questionnaires}
A key feature to conduct in-the-wild studies is the ability to prompt users to answer and fill in short questionnaires and scales. As such, the Wildkey toolkit has the ability to schedule and prompt users to answer custom made questionnaires. The toolkit currently supports the creation of questionnaires with: slider scales, button scales, multiple choice list, single choice, time of day answers and  open end questions (\ref{fig:fig2}). The questionnaires were developed to be easily extendable with new types of questions/answers. Questions are provided one at a time in the sequence defined by the developer/researcher on the Study Management app. To enable users to control the flow of the questionnaire we relied on the library SurveyKit \cite{quickbir94:online} to enable back/cancel functionalities. Answers are collected and synchronized with the cloud database.

\subsubsection{Custom Made Tasks}
Similarly to questionnaires, Wildkey supports the creation of custom made tasks to schedule and deploy to its users. Currently, the toolkit has one example of such a task: the Alternate Finger Tapping test \cite{Lee2016}. 

\subsection{Study Management}
The React app has two key features: study management; and analytics. The application is currently deployable with local hosting and requires user authentication to access, create and edit studies. The study management is divided into User Studies, Tasks, and Users. In User Studies one can check and manage all the existing study schedules and create new study schedules based on the list of available tasks. In Tasks developers/researchers check existing tasks, and can create new tasks of the types described before (i.e. transcriptions, compositions, questionnaires or custom tasks). Lastly, in Users, each user is represented by a tokenID which can be associated with a study and with a schedule.  

\subsubsection{Creating Tasks}
Through the study manager one can create four types of tasks. To create \textbf{1) transcription} or \textbf{2) compositions} tasks you can add the sequence of sentences/questions that you wish the user to do, you can add a timeout for the task (e.g. after 2 minutes the task does not ask for any additional sentences to be transcribed) and you can activate a randomizer of the order by which the sentences/questions are shown.  To create \textbf{3) questionnaires}, first you have to create individual questions by selecting from the types available (i.e.  slider scales, button scales, multiple choice list, single choice, time of day answers and  open end questions) and filling in all required information (e.g. title, description, scales labels). Next, you can create a questionnaire by selecting all the questions that compose it; each question can be flagged as required. Lastly, for \textbf{4) Custom Made} tasks, the configurable parameters will depend on the task.

\begin{figure}
 \includegraphics[scale=0.6]{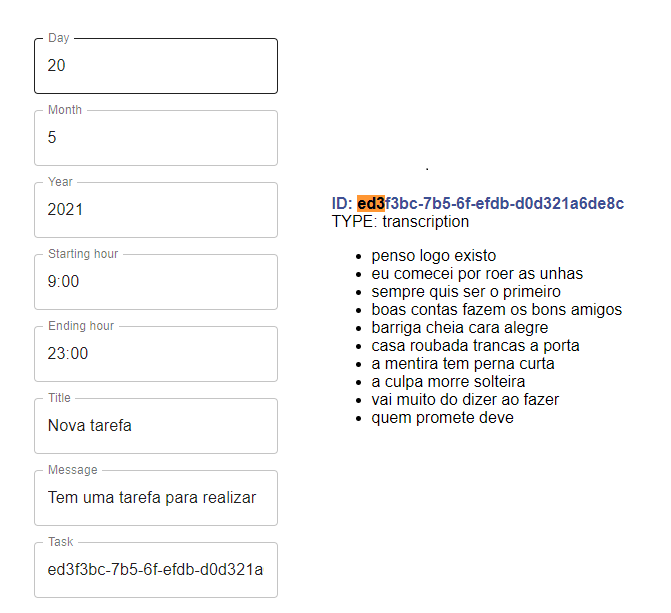}
 	\centering
  \caption{User Management App. Transcription Task configuration.}
  \label{fig:fig3}
\end{figure}

\subsubsection{Creating a User Study }
Through the React app one can create a user study that is composed of a title, start and end date, and one or more \textbf{schedules}. In turn, a schedule is composed of: a start and end date-time and a set of \textbf{timeframes}. A timeframe is an interval of time within a day in which we want to make one or more tasks available. Each timeframe is composed of a start and end time of the interval, a title and description for the notification to be shown to users, and a task list which is composed of the tasks we wish to add for that particular interval (that we have created before, in tasks). An example is shown in Figure \ref{fig:fig3} where on the 20th of May between 9:00 and 23:00 the user could do the transcription task that is composed of the 10 sentences shown. When a schedule has been created, it can be edited, duplicated, deleted;  it also has quick scheduling functions to allow shifting the scheduling by one day or week at a time to facilitate deploying multiple schedules where different users are onboarded on different days. 

\subsubsection{Study Dashboard}
Through User Studies one can check all the schedules created and which users are attributed to which schedules. Each study has a \textbf{dashboard} that allows developers/researchers to quickly check the text-entry metrics collected by task type (e.g. average Words per Minute, Characters Written in transcriptions) associated with each user, and query for the metrics over specific period (e.g. Characters written in the last two days). The dashboard enables researchers/developers to quickly oversee the participants compliance and overall performance. As with the tasks, the Study Dashboard was developed to be easily expandable to accommodate other metrics which may be fundamental to the studies you desire to deploy. 

\subsubsection{User List}
The user list is composed of token ids which represent the users which installed the keyboard application and have registered within the keyboard app. Each user can be associated with a study, and with a schedule within that study. Moreover, one can check a detailed view of the data that participants contributed and overall metrics (e.g. written characters). Similar to the study dashboard, one can also query for data over a specific period. Lastly, one can make a special request to download all data produced by a single user and delete all of it if requested.

\begin{table}
    \centering
    \begin{tabular}{lcccc}
    \hline
        \textbf{Data} & \textbf{Implicit} & \textbf{Transcription} & \textbf{Composition} & \textbf{Demo} \\ \hline
        \textbf{Speed}  \\ \hline
        Words per Minute & \checkmark & \checkmark & \checkmark & \checkmark \\ \hline
        Time per Word & \checkmark & \checkmark & \checkmark & \checkmark \\ \hline
        \textbf{Errors} \\ \hline
        Corrected Error Rate & \checkmark & \checkmark & \checkmark & \checkmark \\ \hline
        Uncorrected Error Rate & \checkmark & \checkmark & \checkmark & \checkmark \\ \hline
        Total Error Rate & \checkmark & \checkmark & \checkmark & \checkmark \\ \hline
        Insertion Error Rate &  & \checkmark & \checkmark & \checkmark \\ \hline
        Omission Error Rate &  & \checkmark & \checkmark & \checkmark \\ \hline
        Substitution Error Rate &  & \checkmark & \checkmark & \checkmark \\ \hline
        Error Correction Attempts & \checkmark & \checkmark & \checkmark & \checkmark \\ \hline
        \textbf{Touch Dynamics }  \\ \hline
        Flight Time & \checkmark & \checkmark & \checkmark & \checkmark \\ \hline
        Hold Time & \checkmark & \checkmark & \checkmark & \checkmark \\ \hline
        Touch Major/Minor & \checkmark & \checkmark & \checkmark & \checkmark \\ \hline
        Touch Offset & \checkmark & \checkmark & \checkmark & \checkmark \\ \hline
        Key Selected &  & \checkmark & \checkmark &  \\ \hline
        Motion Info &  & \checkmark & \checkmark &  \\ \hline
        Timestamp & \checkmark & \checkmark & \checkmark & \checkmark \\ \hline
       \textbf{ Action and Character Counts}  \\ \hline
        Action Count & \checkmark & \checkmark & \checkmark & \checkmark \\ \hline
        Correction Action Count & \checkmark & \checkmark & \checkmark & \checkmark \\ \hline
        Entry Action Count & \checkmark & \checkmark & \checkmark & \checkmark \\ \hline
        Number of Auto Corrects & \checkmark & \checkmark & \checkmark & \checkmark \\ \hline
        Number of Changed Characters & \checkmark & \checkmark & \checkmark & \checkmark \\ \hline
        Number of Selected Suggestions & \checkmark & \checkmark & \checkmark & \checkmark \\ \hline
        Number of Written Characters & \checkmark & \checkmark & \checkmark & \checkmark \\ \hline
        Number of Written Numbers & \checkmark & \checkmark & \checkmark & \checkmark \\ \hline
        Number of Written Special Characters & \checkmark & \checkmark & \checkmark & \checkmark \\ \hline
       \textbf{Other}  \\ \hline
        Raw Text &  & \checkmark & \checkmark & \checkmark \\ \hline
        Intent Calculation & \checkmark & \checkmark & \checkmark & \checkmark \\ \hline
        Intent Validation &  & \checkmark &  &  \\ \hline
        Input Timestamp & \checkmark & \checkmark & \checkmark & \checkmark \\ \hline
        Keyboard Language & \checkmark & \checkmark & \checkmark & \checkmark
\\ \hline
    \end{tabular}
    \caption{\label{tab:table1}List of the Metrics captured by WildKey.}

\end{table}
\subsection{User List}
All metrics collected are calculated locally on the device before being sent out to the cloud database. By default, all data is stored in a JSON format and is readily available for download to post-process it wherever you desire. In Table \ref{tab:table1} you can find all the metrics collected and check in which of the types of data collection WildKey is able to calculate each of the metrics. We divided the collected and calculated metrics into five types: speed, errors, touch dynamics, action and character counts and other. The metrics shown in Table \ref{tab:table1} are only the ones associated with text-entry. All questionnaire answers produce data which is also stored in the same database. Below, we provide a brief description of all the text-entry metrics collected.

\subsubsection{Speed}
Speed metrics are related to how fast a user is writing. We are able to calculate speed in all types of tasks. Speed is calculated taking into account the time from the first entered character to the time of the last character entered. Whenever the keyboard is closed in implicit data collection, we consider the text-entry session to be completed and calculate the speed accordingly. In composition and transcription tasks, only when the user confirms the submission of the task, Wildkey computes speed and all other metrics. 

\begin{itemize}
\item \textbf{Words per minute. }Calculated by (length of transcribed text )/(five characters per word) x (60 seconds/trial time in seconds) \cite{nicolau2017}.
\item \textbf{Time per word.} Calculated by (number of words)/(trial time in seconds)
\end{itemize}

\subsubsection{Errors}
Error rate metrics are an approximation, and a characterization of the errors users made while writing. In transcription tasks, we provide the user with the target phrase which enables us to calculate error rates without any caveats. In implicit collection and in composition tasks, WildKey relies on the spell checker to predict the user intent, as such one can always expect higher error rates due to abbreviations, use of multiple letters in a  row to emphasize text (e.g. noooo) among others, all which will be potentially identified as errors (the process is previously detailed for compositions in the explicit text-entry collection section). 

Wildkey calculates error rates for all types of tasks, but not all are available during implicit data collection due to some requiring significant computational power for longer writing sessions. Below we briefly detail all error rate metrics.

\begin{itemize}
\item \textbf{Corrected Error Rate.} Of the characters erased, the percentage that was erroneous. See \cite{nicolau2017}.
\item \textbf{Uncorrected Error Rate.} Percentage of erroneous characters in the final transcribed sentence \cite{nicolau2017,Wobbrock2006}.
\item \textbf{Total Error Rate. } Of the characters entered, the percentage that was erroneous, corrected, or not \cite{nicolau2017}.
\item \textbf{Insertion Error Rate.} Additional erroneous characters added (corrected and uncorrected). We can provide both individually by letter and aggregated measures. See \cite{nicolau2017,Wobbrock2006}.
\item \textbf{Omission Error Rate.} Corrected (characters that were missing at first but were backspaced and inserted) and Uncorrected Omitted characters in relation to the number of times the character was presented. We can provide both individually by letter and aggregated measures. See \cite{nicolau2017,Wobbrock2006}.
\item \textbf{Substitution Error Rate. } Ratio of substitutions to intentions. We can provide both individually by letter and aggregated measures. See \cite{nicolau2017,Wobbrock2006}. 
\item \textbf{Error Correction Attempts.}	Number of corrections sequences, meaning how many times the user started a sequence of correction actions (i.e. backspaces). 

\end{itemize}

\subsubsection{Touch Dynamics}
WildKey collects a variety of touch related metrics (Table \ref{tab:table1}). For explicit tasks, all touch metrics are calculated since the text content the user writes is prompted by Wildkey and raw text content is collected. For implicit collection, no touch metrics are collected that would enable anyone to reconstruct the text written (i.e. keys selected, touch points).
\begin{itemize}
\item \textbf{Flight time}. Sequence of values corresponding to the time between the release of two key taps \cite{gallego2017}.
\item \textbf{Hold Time}.  Sequence of values of time spent touching the screen in each touch.
\item \textbf{Touch Major/Minor}. Sequence of  values of the TouchMajor (length of the major axis of an ellipse that represents the touch area) and the TouchMinor (length of the major axis of an ellipse that represents the touch area).
\item \textbf{Touch offset}. Sequence of values of the differences in key centroids and hitpoint deviations (i.e., x and y offsets of touch gestures with regard to individual keys) \cite{Guerreiro2015}, measured in pixels, and converted to cm given the device specifics. 
\item \textbf{Key selected}. The sequence of keypresses.
\item \textbf{Motion info}. Sequence of all the touch motions detected by Android and their timestamp (i.e. down, move, up). 

\end{itemize}

\subsubsection{Action and Character Counts }
To enable us to better characterize the users text-entry behaviours, Wildkey collects a wide variety of action and character counts.
\begin{itemize}
\item \textbf{Action Count}. Total number of actions performed (i.e. corrections and entry actions). 
\item \textbf{Correction Action Count}. Total number of individual actions that corrected input (i.e. substitutions and deletions).
\item \textbf{Entry Action Count}. Total number of individual actions that produced an input. Using a suggestion counts as a single entry action. 
\item \textbf{Number of Auto Corrects}. Total number of auto-corrects.
\item \textbf{Number of Changed Characters}. Total number of changes to the input stream.	
\item \textbf{Number of Selected Suggestions}.	Number of times suggestions were selected.
\item \textbf{Number of Written Characters}. Total number of written characters.	
\item \textbf{Number of Written Numbers}. Total number of written numbers.	
\item \textbf{Number of Written Special Characters}. Total number of written special characters.
\end{itemize}

\subsubsection{Other}
In other we aggregate the data/metrics that are related with intent, raw text and other device/system characteristics. 
\begin{itemize}
\item \textbf{Raw Text. }In tasks where we explicitly prompt the user we collect the input stream of all user interactions with the keyboard.
\item \textbf{Intent Calculation.} When we do not provide the target sentence for the user to transcribe, Wildkey estimates the user intent. The final target phrase is calculated and available for every text-entry session.
\item \textbf{Intent Validation.} In transcription tasks, we also calculate the user's final target phrase through intent calculation. Although we rely on the target phrase we provide to generate all metrics, the calculation of the intent allows us to have a baseline of the intent calculation performance which we refer to as intent validation. 
\item \textbf{Input Timestamp.} A sequence of all touchdown input timestamps, including selecting suggestions and making cursor changes.
\item \textbf{Keyboard Language. }For every session we record the current language of the keyboard and any changes during the text-entry session.

\end{itemize}

\subsection{Privacy-Aware}
Smartphones are becoming an extension of oneself \cite{Park2019} and data privacy has become of the utmost importance. Text content can be one of the most private types of data users generate and approaches that seek to collect this type of information can expect to be met with resistance by some users and overall suffer from lower levels of adherence and compliance. Aware of the challenges WildKey was devised to not store any textual content outside the specific tasks it asks users to perform. Moreover, all metrics that would enable the reconstruction of the text content are not available when analysing users text input behaviours implicitly. WildKey calculates all metrics locally on the device and only stores the results of those calculations in the cloud. Furthermore, when the user is inserting only numbers or in passwords fields no analysis is conducted. To ensure users are in control of the data they are sharing, WildKey has an always available button on the top left of the keyboard to activate an “incognito mode”, which resets every time the keyboard is closed. When active no data is analysed. Lastly, as with other Android keyboards, users can quickly swap to another active keyboard by holding the space bar for a quick options menu.

WildKey currently relies on Firebase services for its cloud storage which encrypts data in transit using HTTPs and logically isolates customer data. All users data is pseudo-anonymised within the database where users are only identifiable by their tokenID. Cross reference between tokenIDs and users identification is not stored in the same database. 

To facilitate GDPR compliance in regards to the ‘right to be forgotten’ and the right users have to request their data, in the Study Management application under users, WildKey enables researchers to quickly download all data from a specific user, and delete it all if necessary.

Lastly, in the github repository for the WildKey React app we provide a mock up report (pdf file) which shows the type of data WildKey is able to collect in a comprehensive manner. The report can be leveraged by researcher/developers to enable users to make a conscious choice when deciding to use the keyboard or participate in a user study.

\subsection{Limitations}
WildKey is currently only available on Android, supporting devices with operating systems 6.0 and above which covers about 84.9\% of all devices according to Android Platform - API Version Distribution. The number of supported devices will continue to grow as deprecated devices cease to function and get replaced. 

For composition and implicit text collection we rely on intent prediction to calculate error related metrics which artificially increases the error rates due to a variety of text-entry behaviours people have in informal conversations (e.g. abbreviations). Moreover, the intent prediction is only as accurate as the used dictionary and spell checker. For words that exist in the dictionary, all entered are considered correct, but for words that are not detected we use the spell checker best prediction, which may vary across different spell checkers and dictionaries. We relied on the open source Android OS dictionary and spell checker which provides support for 26 languages. The quality of the intent prediction can additionally vary among different languages. While WildKey currently does not provide a streamline process to change or add dictionaries and/or spell checkers it is possible for developers to customize these options. 

Although the toolkit was designed to support a wide variety of in-the-wild study protocols it may require additional development efforts for custom-tailored deployments. The demo available on the Playstore provides a standalone version with local storage which facilitates deployment, but complicates data retrieval and study oversight. To leverage the full potential of the toolkit developers/researchers have to deploy the WildKey toolkit ecosystem which currently requires some degree of technical expertise.

\section{Conclusions}
The data one can collect from typing sessions has continually shown its potential for a variety of different domains. As a complex task that combines motor and cognitive functions, the features one can extract can be explored for disease detection and monitoring, for biometrics, personalized assistive technology, and within the quantified self movement. To support and promote further explorations we developed and made available to all the  WildKey Toolkit with its first release. The toolkit is currently being leveraged in multiple studies with topics from privacy and compliance, to fatigue, to monitoring motor fluctuations in neurodegenerative diseases. The toolkit is an ongoing open source project and we welcome contributions. New features will continue to be added as new requisites appear to support studies with different goals and in different contexts. 

We are very keen to hear about your work. Please reach out if you decide to use the toolkit, have any questions or desires for additional features

\section{Acknowledgements}
We would like to thank our collaborators from OpenLab Newcastle as the first external team to use the keyboard toolkit providing us with valuable feedback and contributions to the project. This project was partially supported by  FCT through LASIGE Research Unit funding, ref. UIDB/00408/2020 and ref. UIDP/00408/2020, and by the IDEA-FAST project which has received funding from the Innovative Medicines Initiative 2 Joint Undertaking under grant agreement No. 853981. This Joint Undertaking receives support from the European Union’s Horizon 2020 research and innovation programme and EFPIA and associated partner.

\bibliographystyle{unsrtnat}
\bibliography{references}  %%% Uncomment this line and comment out the ``thebibliography'' section below to use the external .bib file (using bibtex) .





% \end{thebibliography}

\end{document}